\documentclass[aps, prb,  twocolumn, pt=10, showpacs]{revtex4-1}

\usepackage{graphicx}
\usepackage{color}
\usepackage{amsmath}
\usepackage{amssymb}
\usepackage{amsthm}
\usepackage{times}
\usepackage{hyperref}

\newcommand{\mathsym}[1]{{}}

\def  \bsig    {\mbox{\boldmath$\sigma$}}

\def  \btau    {\mbox{\boldmath$\tau$}}

\begin{document}
\draft
\title{Current-induced spin polarization and spin-orbit torque in graphene}
\author{A. Dyrda\l$^{1}$ and J.~Barna\'s$^{1,2}$}
\address{$^1$Faculty of Physics, Adam Mickiewicz University,
ul. Umultowska 85, 61-614 Pozna\'n, Poland \\
$^2$
Institute of Molecular Physics, Polish Academy of Sciences,
ul. M. Smoluchowskiego 17, 60-179 Pozna\'n, Poland
}

\date{\today }

\begin{abstract}
Using the Green function formalism we calculate a current-induced spin polarization of weakly magnetized graphene with Rashba spin-orbit interaction. In a general case, all components of the current-induced spin polarization are nonzero, contrary to the nonmagnetic limit, where the only nonvanishing component of spin polarization is that in the graphene plane and normal to electric field. When the induced spin polarization  is exchange-coupled to the magnetization, it exerts a spin-orbit torque on the latter. Using the Green function method we have derived some analytical formulas for the spin polarization and also determined the corresponding spin-orbit torque components. The analytical results are compared with those obtained numerically. Vertex corrections due to scattering on randomly distributed impurities is also calculated and shown to enhance the spin polarization calculated in the {\it bare bubble} approximation.
\end{abstract}

\maketitle

\section{Introduction}

One of the main issues of the present-day spin electronics, that is of great importance for further development of high-density memory devices and magnetic random access memories, is effective  manipulation of magnetization by a spin-polarized current. The general idea of switching and controlling orientation of a magnetic moment with electric current flowing through a system is based on coupling between the electron spin and magnetic moments. Two such interactions turned out to be especially useful -- exchange interaction and spin-orbit coupling.

In a magnetically nonuniform system, the spin-polarized current generates a torque that is a consequence of: (i) exchange coupling between the conduction electrons and magnetization, and (ii) conservation of angular momentum in the system. The torque appears then as a result of the spin angular momentum transfer from a spin-polarized current (or pure spin current) to magnetic moments. Therefore, this torque is called spin-transfer torque.~\cite{Manchon2007,Ralph2008} Such a torque leads, among others, to magnetic switching in spin vales and to domain wall displacements, as observed recently in many experiments. Moreover, these phenomena give a possibility to construct low-power non-volatile memory cells (STT-MRAM, racetrack memory), integrated circuits employing a logic-in-memory architecture as well as logic schemes processing information with spins.~\cite{Brataas2012}

Another possibility to control orientation of magnetic moments is based on a spin torque that appears due to spin-orbit interaction in the system. The corresponding torque exerted on the magnetization is usually referred to as the spin-orbit torque, and appears also in a magnetically uniform system, like a single uniform layer.  Physical mechanism of the spin-orbit torque is based on a nonequilibrium spin polarization of the system, which is induced by an external electric field (current) in the presence of spin-orbit interaction. Such a spin polarization was predicted long time ago in nonmagnetic systems, where an electric current flowing through the system with spin-orbit interaction was shown to induce not only the transverse spin-current~\cite{dyakonov71_she,hirsch} (so-called spin Hall effect), but also a spin-polarization of conduction electrons.~\cite{dyakonov71,aronov89,edelstein90,aronov91,Golub,Shen} In the case of two-dimensional electron gas with Rashba spin-orbit interaction, the induced spin polarization is in the plane of the electron gas and normal to the electric field.
Such a nonequilibrium spin-polarization may be treated as an effective magnetic field, which may lead to  reorientation of a magnetic moment, and also can modify or induce magnetic dynamics. The  spin-orbit torque was analyzed in recent few years in many papers, mainly in metallic and  semiconductor heterostructures.~\cite{Zhang2008,Zhang2009,Abiague2009,Manchon,Gambardella,Garello2013} While the current-induced spin polarization, known also as the inverse spin-galvanic effect,~\cite{Ganichev2002} is well known and was investigated theoretically as well as experimentally in the recent three decades, the role of geometric phase in this effect, and consequently in the spin-orbit torque, was invoked only very recently.~\cite{Kurebayashi,Avci,Li2014}

In this paper we consider the current-induced spin polarization and spin-orbit torque in graphene, which is assumed to be  deposited on a substrate that ensures the presence of spin-orbit interaction of Rashba type~\cite{Dedkov2008}. We also assume that the graphene is magnetized, which may be either due to the magnetic proximity effect to a ferromagnetic substrate (or cover layer), or  due to magnetic atoms (nanoparticles) on its surface.~\cite{Yokoyama2008,Haugen2008,ZPNiu2011,ZPNiu2014,KawakamiZutic2014}
Coexistence of  the Rashba spin-orbit interaction and proximity-induced magnetism in graphene was predicted theoretically and also observed experimentally.\cite{KawakamiZutic2014,Qiao2010,Qiao2014,Wang2015,Gong2011}
As the spin transfer torque in ferromagnetic graphene junctions  was already considered theoretically (see e.g. Yokoyama and Linder~\cite{Yokoyama}),  the problem of spin torques induced by spin-orbit interaction in graphene is rather unexplored.

It has been shown that the current-induced spin polarization in a defect-free nonmagnetic graphene with Rashba spin-orbit interaction is oriented in the graphene plane and is also normal to the current orientation. Moreover, sign of the spin-polarization depends on the chemical potential and also on the sign of the Rashba spin-orbit coupling parameter.~\cite{Dyrdal2014} When the Fermi level passes through the Dirac points, the spin polarization becomes reversed. In this paper we show that the  current-induced spin polarization in magnetic graphene  has generally all three components. In the approximation linear with respect to the magnetization, one of these components is equal to that in the case of a nonmagnetic graphene, i.e. it is proportional to the relaxation time. The leading terms in the other two components are independent of the relaxation time.

The paper is organized as follows. In Sec. 2 we describe the model and present a general formula describing current-induced spin polarization. Analytical formulas as well as numerical results for the current-induced spin polarization are presented in  Sec. 3. Vertex correction is calculated in section 4, while the spin-orbit torque is
described and discussed in Sec. 5. Summary and final conclusions are in Sec. 6.

\section{Model and method}

Transport properties of graphene close to the charge neutrality point are determined mainly by electrons in the vicinity of Dirac points.
The corresponding effective-mass Hamiltonian, $H^0_K$, which describes the low-energy electronic states in graphene around the $K$ point of the Brillouin zone,
can be written as a sum of three terms,~\cite{kane}
\begin{equation}
H^{0}_{K} = H_{0} + H_{R} + H_{\bf{M}} .
\end{equation}
The first term,  $H_{0}$, describes the low energy electronic states of pristine graphene, and can be written as a matrix in the pseudospin (sublattice) space,
\begin{eqnarray}
\label{hk0}
H_{0} = v \left(
      \begin{array}{cc}
        0 & (k_{x} - i k_{y}) \sigma_{0} \\
        (k_{x} + i k_{y}) \sigma_{0} & 0 \\
      \end{array}
    \right),
\end{eqnarray}
where  $v = \hbar v_{F}$, with $v_{F}$ denoting the electron velocity in graphene, which is constant.
The second term in Eq.~(1) describes the Rashba spin-orbit interaction due to a substrate,
\begin{eqnarray}
\label{hkR}
H_{R} =  \lambda \left(
                \begin{array}{cc}
                  0 & \sigma_{y} + i \sigma_{x} \\
                  \sigma_{y} - i \sigma_{x}  & 0 \\
                \end{array}
              \right),
\end{eqnarray}
 with $\lambda$ being the Rashba spin-orbit coupling parameter. The last term of the Hamiltonian (1) represents the influence of an effective exchange field $\tilde{\mathbf{M}}$ created by a nonzero magnetization. Such a magnetization can appear in graphene, for instance, due to the proximity effect to a magnetic substrate. This term can be written in the form,
  \begin{eqnarray}
\label{hkM}
H_{\bf{M}} =  -  \tilde{\mathbf{M}} \cdot  \left(
                \begin{array}{cc}
                  \bsig & 0 \\
                  0  & \bsig\\
                \end{array}
              \right),
\end{eqnarray}
 where the exchange field $\tilde{\mathbf{M}}$ is measured in energy units. This field can be related to the magnetization $\mathbf{M}$ and the local exchange interaction between the conduction electrons and  magnetization in the two-dimensional graphene, $J_{\rm ex}({\bf r}-{\bf r}^\prime )= J_{\rm ex}\,\delta ({\bf r}-{\bf r}^\prime )$, {\it via} the formula $\tilde{\mathbf{M}} = (J_{\rm ex}/2g\mu_B){\mathbf M}$, Here,
$g$ is the Lande factor ($g=2$), $\mu_B$ is the Bohr magneton, while positive and negative $J_{\rm ex}$ (measured in the units of Jm$^2$) correspond to antiferromagnetic and ferromagnetic coupling, respectively.
 In the above equations, $\bsig$ is the vector of Pauli matrices, $\bsig =(\sigma_x,\sigma_y, \sigma_z )$, while the matrix $\sigma_{0}$ denotes the unit matrix in the spin space. Note that the so-called intrinsic spin-orbit interaction in graphene is very small and therefore it is neglected in our consideration. In a general case, the magnetization vector $\mathbf{M}$ may be oriented arbitrarily in space, and its orientation will be described by two spherical angles, $\theta$ and $\xi$, as indicated in Fig.~1. Moreover, the absolute magnitude of $\bf M$ is assumed to be constant, $|{\bf M} |\equiv M={\rm const}$. Hamiltonian for the second non-equivalent Dirac point, $K^\prime$, can be obtained from $H_K$ by reversing sign of the wavevector component $k_x$ and substitution $\sigma_{y} \rightarrow - \sigma_{y}$ in $H_{R}$.

\begin{figure}[t]
  \centering
  \includegraphics[width=220pt]{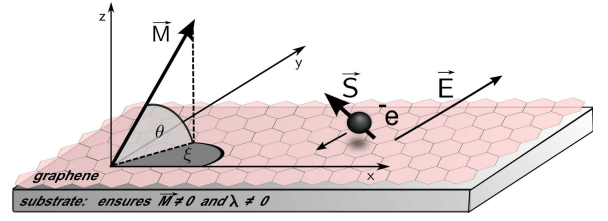}\\
  \caption{(Color online) Schematic of the system under consideration. Graphene is on a substrate which assures a nonzero magnetization and also a spin-orbit interaction of Rashba type. Orientation of the magnetic moment $\bf M$ is described by the angles $\theta$ and $\xi$. An external electric field is oriented along the axis $y$.}
  \label{schematic}
\end{figure}

In the lowest order with respect to the exchange field $\tilde{\mathbf{M}}$, the casual Green function corresponding to the Hamiltonian (1), $G_{k}^{0} = \{ [\varepsilon + \mu + i \delta\, {\rm sign}\,(\varepsilon)] - H^{0}_{K}\}^{-1}$ has poles at $\varepsilon = E_{n} - \mu - i \delta\,{\rm sign}\,(\varepsilon)$, where $E_{n}$ ($n = 1-4$) are eigenvalues of the Hamiltonian (1) without the term $H_{\mathbf{M}}$. These eigenvalues have the following form:
\begin{eqnarray}
E_{1,2} = \mp \lambda - \sqrt{k^{2}v^{2} + \lambda^{2}}\\
E_{3,4} = \mp \lambda + \sqrt{k^{2}v^{2} + \lambda^{2}},
\end{eqnarray}
where $E_{1,2}$ correspond to the valance bands, while $E_{3,4}$ describe the conduction bands. Note, the bands corresponding to $n=2$ and $n=3$ touch each other at the Dirac point ($k=0$), while a gap equal to $4\lambda$ appears between the bands $n=1$ and $n=4$.

In the presence of a dynamical external electric field applied along the axis $y$, the total Hamiltonian for electrons  near the $K$ point takes the form
\begin{equation}
H = H^{0}_{K} + H_{K}^{\mathbf{A}},
 \end{equation}
where the second term,
 \begin{equation}
 H_{K}^{\mathbf{A}} = - e \hat{v}_{y} A_{y}(t) = - i e \frac{v}{\hbar}  \left( \begin{array}{cc}
        0 & -\sigma_{0} \\
        \sigma_{0} & 0 \\
      \end{array}\right) A_{y}(t),
 \end{equation}
is the perturbation due to interaction with the time-dependent electromagnetic field represented by the vector potential $A_{y}(t) = A_{y}e^{-i \omega t}$. Here, $e$ is the electron charge, $\hat{v}_y$ is the $y$-component of the electron velocity operator, $\hat{\mathbf v}=\partial H^0_K/\partial \mathbf{k}$, whereas $\omega$ is the frequency of the dynamical field (later we will take the limit of $\omega \to 0$).

When an electric current flows in the system due to the electric field, electron spins  become polarized as a result of the co-operation of the current and Rashba spin-orbit coupling. This nonequilibrium spin polarization of conduction electrons can be calculated (in the zero-temperature limit) using the following formula:
\begin{equation}
\label{S_alpha}
S_{\alpha}(t) = - i \mathrm{Tr}\int \frac{d^{2} \mathbf{k}}{(2 \pi)^{2}} \hat{S}_{\alpha} G_{\mathbf{k}}(t, t^\prime)|_{t^\prime =t+0},
\end{equation}
where $G_{\mathbf{k}}(t, t^\prime)$ is the zero-temperature causal Green function corresponding to the total Hamiltonian $H$ (see Eq.~7), and $\hat{S}_{\alpha}$ is the spin vertex function defined as
\begin{equation}
\label{sa}
\hat{S}_{\alpha} = \frac{\hbar}{2} \left( \begin{array}{cc}
        \sigma_{\alpha} & 0 \\
        0 & \sigma_{\alpha} \\
      \end{array}\right).
\end{equation}
Upon Fourier transformation with respect to the time variables and expansion in a series with respect to the vector potential $A_{y} = - i E_{y}/\omega$,
the expression (\ref{S_alpha}) for the induced nonequilibrium spin density takes the form
\begin{eqnarray}
\label{S_op}
S_{\alpha} (\omega) = \frac{e E_y}{ \omega} {\rm{Tr}}\int\frac{d^{2}\mathbf{k}}{(2\pi)^{2}} \int\frac{d\varepsilon}{2 \pi}
\hat{S}_{\alpha}
         \;G^0_{\mathbf{k}}(\varepsilon + \hbar\omega) \hat{v}_{y} G^0_{\mathbf{k}}(\varepsilon).\hspace{0.6cm}
\end{eqnarray}
In the dc limit, $\omega \rightarrow 0$, the above formula leads to the following expression for the spin polarization:
\begin{eqnarray}
\label{S_alpha1}
S_{\alpha} = \frac{e}{2 \pi} E_{y} \hbar {\rm{Tr}}\int\frac{d^{2}\mathbf{k}}{(2\pi)^{2}}\hat{S}_{\alpha}  G^{0R}_{\mathbf{k}} \hat{v}_{y} G^{0A}_{\mathbf{k}},
\end{eqnarray}
where $G^{0R(A)}_{\mathbf{k}}$ is the retarded (advanced) Green function corresponding to the unperturbed Hamiltonian (1), taken at the Fermi level ($\varepsilon = 0$).
Upon taking into account Eqs~(\ref{sa}) and (\ref{S_alpha1}), and also including the contribution from the second Dirac  point, the expression for the induced spin polarization acquires the form
\begin{eqnarray}
\label{S_alpha_main}
S_{\alpha} = \frac{e\hbar^{2}}{2 \pi} E_{y} {\rm{Tr}}\int\frac{d^{2}\mathbf{k}}{(2\pi)^{2}}\left(
      \begin{array}{cc}
      \sigma_{\alpha} & 0\\
       0 & \sigma_{\alpha} \end{array} \right) G^{0R}_{\mathbf{k}} \hat{v}_{y} G^{0A}_{\mathbf{k}}.
\end{eqnarray}
Based on this formula we calculate analytically as well as numerically the current-induced spin polarization, as described and  discussed in the subsequent section.

\section{Current-induced spin polarization}

From the general formula (\ref{S_alpha_main}) one finds the following expression for the $\alpha$-th component of the spin polarization:
\begin{equation}
\label{12}
S_{\alpha} = \frac{e \hbar}{2 \pi}E_{y} \int \frac{dk k}{(2\pi)^{2}} \frac{T_{\alpha}}{\Pi_{n=1}^{4}(\mu - E_{n} + i \Gamma)(\mu - E_{n} - i \Gamma)},
\end{equation}
where $T_{\alpha}$ is defined as
\begin{equation}
\label{11}
T_{\alpha} = \hbar \int_{0}^{2 \pi} d\phi {\rm{Tr}} \left[ \left(
      \begin{array}{cc}
      \sigma_{\alpha} & 0\\
       0 & \sigma_{\alpha} \end{array} \right) g^{0R}_{\mathbf{k}} \hat{v}_{y} g^{0A}_{\mathbf{k}}\right].
\end{equation}
Here, $g^{0R(A)}_{\mathbf{k}}$ is the nominator of the retarded (advanced) Green function, $\phi$ stands for the angle between the axis $x$ and the wavevector $\bf k$, while $\Gamma = \hbar/2 \tau$, where $\tau$ is the momentum relaxation time. The parameter $\Gamma$ (or equivalently relaxation time $\tau$) will be treated here as a phenomenological parameter, which effectively includes contributions due to momentum relaxation  from various scattering processes (scattering on impurities, other structural defects, phonons, or electron-electron scattering). Note that $\Gamma$ depends in general on the chemical potential $\mu$ and may be also different in the two Rashba subbands. However, when the Fermi level is in the two subbands, we assume for simplicity the same $\Gamma$ for both of them.

Up to the terms  linear in the exchange field $\tilde{M}$, the functions $T_{\alpha}$ ($\alpha =x, y, z$), see Eq.~(15), can be written as follows:
\begin{subequations}
\begin{align}
\label{Ta}
T_{x} = 16 \lambda \pi v \mu (k^{4} v^{4} - \mu^{4} + 4 \lambda^2 \mu^{2}),\\
T_{y} =  64 \pi v \lambda \tilde{M} \mu (k^{2} v^{2} - 2 \lambda^{2}) \Gamma \sin\theta, \label{Tb}\\
T_{z} = - 64 \pi \lambda \tilde{M}  v^{3} k^{2} \mu \Gamma \cos\theta \sin\xi .\label{Tc}
\end{align}
\end{subequations}
Note, the dependence on the orientation of $\bf M$ is contained in the above expressions for $T_y$ and $T_z$, while $T_x$ is independent of $\bf M$.
Equations (14) and (16) allow finding spin polarization in a general case, i.e. for an arbitrary relaxation time. However,
some analytical expressions for all components of the spin polarization can be obtained in the limit of low impurity concentration, i.e. for long relaxation times ($\tau \rightarrow \infty $).

Consider first the $x$-component of the spin polarization.
Combining Eq.(\ref{Ta}) with Eq.(\ref{12}) and making the substitution $\sqrt{k^2 v^2 + \lambda^2} = \gamma$, one obtains
\begin{eqnarray}
\label{Sx_0a}
S_{x} = 8 e \hbar E_{y} \lambda \mu \hspace{6cm}\nonumber\\ \times \int_\lambda^\infty \frac{d \gamma \gamma}{v (2\pi)^2} \frac{ (\gamma^2 - \lambda^2)^2 - \mu^4 + 4 \lambda^2 \mu^2}{[(\mu + \lambda + \gamma)^2 + \Gamma^2] [(\mu - \lambda + \gamma)^2 + \Gamma^2]}\nonumber\\
\times \frac{1}{[(\mu + \lambda - \gamma)^2 + \Gamma^2][(\mu - \lambda - \gamma)^2 + \Gamma^2]}.\hspace{0.5cm}
\end{eqnarray}
From this formula follows that $S_{x}$ is independent of $\tilde{M}$ in the linear approximation with respect to the exchange field.
For long relaxation times we get the same analytical formulas  as those in the case of nonmagnetic graphene,~\cite{Dyrdal2014} i.e.,
\begin{equation}
\label{sx1}
S_{x} = \frac{e}{4 \pi} \frac{2 \lambda \pm \mu}{v(\lambda \pm \mu)} \mu E_{y} \tau
\end{equation}
for the Fermi level lying in the range  $-2 \lambda < \mu < 2 \lambda$,
and
\begin{equation}
\label{sx2}
S_{x} = \pm \frac{e}{4 \pi} \frac{2 \lambda}{v (\mu^{2} - \lambda^{2})} \mu^{2} E_{y} \tau
\end{equation}
for $|\mu| > 2 \lambda$. In both above equations (as well as below), the upper and lower signs correspond to $\mu >0$ and $\mu <0$, respectively.

\begin{figure*}[]
\centering
\includegraphics[width=0.9\textwidth]{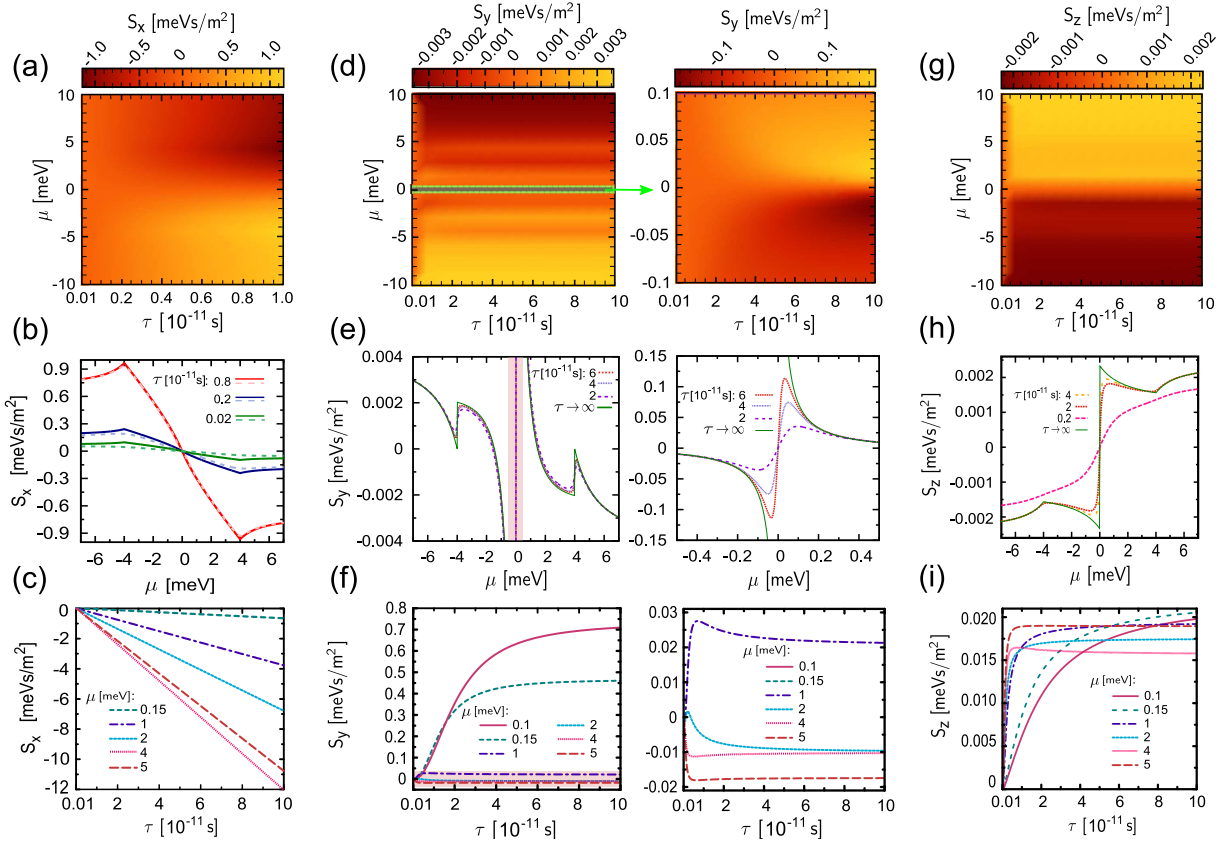}
    \caption{(Color online) Spin polarization components induced by a current: (a,b,c) show the $x$-component, (d,e,f) show the $y$-component, whereas (g,h,i) show the $z$-component. The top panel (a,d,g) shows the spin polarization components as a function of chemical potential $\mu$ and relaxation time $\tau$. The medium panel (b,e,h) shows the spin polarization components as a function of chemical potential $\mu$  for indicated values of $\tau$, while the bottom panel (c,f,i) shows the polarization components as a function of $\tau$ for indicated values of $\mu$.  The right parts of (d,e,f) present the corresponding shaded regions in the left parts. The solid and dashed lines in (b) represent the results based on the analytical formulas and numerical integration, respectively.  The curves for $\tau\to\infty$ in (e,h) correspond to analytical solutions. The other parameters are: $\lambda = 2$ meV,  $E_{y} = 1$ V/cm, $\tilde{M} = 0.1$ meV, $\theta = \pi/3$, and $\xi = \pi/2$.}
    \label{Fig:sx}
  \end{figure*}

The spin polarization given by Eqs (\ref{sx1}) and (\ref{sx2}) is proportional to $\tau$. However, one should bear in mind that these formulas were derived on the assumption of long $\tau$. Therefore, one may expect some deviations from this formula when $\tau$ is finite and not too long.
In Fig.~\ref{Fig:sx}(a) we show variation of the $S_x$ component of spin polarization with the chemical potential $\mu$ and relaxation time $\tau$, obtained by numerical integration of the formula (17). Figures~\ref{Fig:sx}b and \ref{Fig:sx}c, in turn,  present cross-sections of the density plots shown in Fig.~\ref{Fig:sx}(a)  for constant values of $\tau$ and $\mu$, respectively. The results obtained from the analytical formulas are compared in Figs~\ref{Fig:sx}(b) with those obtained by numerical integration of the formula (17). From this comparison follows that for $\tau$  of the order of $10^{-11}$s or smaller, there are some deviations from the results given by the analytical formulas, though these deviations are not large. For $\tau$  of the order of $10^{-10}$s or longer, numerical results match quite well those obtained from the analytical formulas. Since the $S_x$  component is the same in magnetic and nonmagnetic limits (within the approximations used here), and in the nonmagnetic limit it was considered and analyzed in Ref.[\onlinecite{Dyrdal2014}], we will not discuss this component in more detail.

From Eqs (\ref{12}) and (\ref{Tb}) one finds the $y$-component of the spin polarization in the following form:
\begin{eqnarray}
\label{Sy_0}
S_{y} = 32 e E_{y} \hbar \lambda \mu \tilde{M} \hspace{6cm}\nonumber\\ \times \int_\lambda^\infty \frac{d \gamma \gamma}{v (2\pi)^2} \frac{(\gamma^2 - 3 \lambda^2) \Gamma \sin\theta }{[(\mu + \lambda + \gamma)^2 + \Gamma^2] [(\mu - \lambda + \gamma)^2 + \Gamma^2]}\nonumber\\
\times \frac{1}{[(\mu + \lambda - \gamma)^2 + \Gamma^2][(\mu - \lambda - \gamma)^2 + \Gamma^2]}\, .\hspace{0.8cm}
\end{eqnarray}
In the limit of slow relaxation, $\Gamma \to 0$, the above formula leads to the following analytical results:
\begin{equation}
\label{sy1}
S_{y} = \pm \frac{e \hbar}{4\pi} \frac{\tilde{M} }{\lambda} \sin\theta \frac{\mu (\mu \pm 2\lambda) - 2 \lambda^{2}}{2v\mu(\mu \pm \lambda)} E_{y}
\end{equation}
for $|\mu| < 2 \lambda$, and
\begin{equation}
\label{sy2}
S_{y} = \pm \frac{e \hbar}{4\pi} \frac{\tilde{M} }{\lambda} \sin\theta \frac{\mu^{2} - 4 \lambda^{2}}{v(\mu^{2} - \lambda^{2})} E_{y}
\end{equation}
for $|\mu| > 2 \lambda$.

Numerical results for the $y$-component of the current-induced spin polarization, obtained by numerical integration of the formula (20) are shown in Fig.2(d) as a function of chemical potential $\mu$ and relaxation time $\tau$. Figures 2(e,f) present cross-sections of Fig.2(d). Figure~2(e) additionally shows the results obtained from analytical formulas, see the curves for $\tau\to\infty$. Right parts of Fig.2(d,e,f) present in more detail the corresponding shaded regions. Similarly as the $x$-component, $S_y$ is antisymmetric with respect to reversal of the sign of Fermi energy, and its dependence on $\mu$ also reveals some steps at $\mu = \pm 2\lambda$. These steps are associated with the edges of the bands $E_1$ and $E_4$. Moreover, when the Fermi level is at the Dirac point ($\mu = 0$), the analytical solution (\ref{sy1}) for $S_{y}$ becomes divergent.
To understand origin of the divergency in the analytical solution for $\tau\to \infty$,  one should note that the solution for the $x$-component is  also infinite for $\tau\to\infty$, independently of $\mu$. This clearly shows that the limit of $\tau\to \infty$  is not physical as the dissipation processes are necessary in order to stabilize a finite current-induced deviation of the system from equilibrium, and thus also a finite current density and spin polarization. Therefore, in Fig.2(e) we compare  the numerical results based on the corresponding analytical formulas with those obtained by numerical integration.  This comparison clearly shows that the results obtained from the analytical formulas are roughly in agreement with those obtained from numerical integration, except  the vicinity of  $\mu=0$, where the analytical solution diverges for $\mu\to 0$, while the numerical results based on Eq.(\ref{Sy_0}) are then finite. Moreover, some discrepancy also occurs around $\mu=\pm 2\lambda$, but now the difference is finite and rather small.
Thus, one should bear in mind that the analytical results (\ref{sy1}) and (\ref{sy2}) for the $y$-component have limited applicability range, and are not applicable for $\mu$ in the vicinity of the Dirac points.

The $S_{z}$ component can by found from Eqs (\ref{12}) and (\ref{Tc}) and acquires the form
\begin{eqnarray}
\label{Sz_0}
S_{z}= - 32 e E_{y} \hbar \lambda \mu \tilde{M} \cos\theta \sin\xi \hspace{4cm}\nonumber\\ \times \int_\lambda^\infty \frac{d \gamma \gamma}{v (2\pi)^2} \frac{(\gamma^2 - \lambda^2) \Gamma }{[(\mu + \lambda + \gamma)^2 + \Gamma^2] [(\mu - \lambda + \gamma)^2 + \Gamma^2]}\nonumber\\
\times \frac{1}{[(\mu + \lambda - \gamma)^2 + \Gamma^2][(\mu - \lambda - \gamma)^2 + \Gamma^2]}\, .\hspace{0.8cm}
\end{eqnarray}
Similar calculations as those done for the $y$-component lead to the following analytical expressions in the limit of long relaxation time:
\begin{eqnarray}
\label{sz1}
S_{z} = \mp \frac{e \hbar}{4\pi} {\tilde{M}}{\lambda} \cos\theta \sin\xi \frac{\mu \pm 2 \lambda}{2 v (\mu \pm \lambda)} E_{y}
\end{eqnarray}
for $\mu <  2\lambda$, and
\begin{eqnarray}
\label{sz2}
S_{z} = \mp \frac{e \hbar}{4\pi} \frac{\tilde{M}}{\lambda} \cos\theta \sin\xi \frac{\mu^{2} - 2 \lambda^{2}}{v (\mu^{2} - \lambda^{2})} E_{y}
\end{eqnarray}
for $\mu > 2\lambda$.

In Fig.2(g) we present the $z$-component of the current-induced spin polarization, calculated  as a function of the chemical potential and relaxation time by numerical integration of the formula (25). In turn, Figs.2(h,i) show cross-section of Fig.2(g). In Fig.2(h) we additionally compare the numerical results with those obtained from analytical solution. Now, the analytical solution is not divergent, see the curve for $\tau\to\infty$. When the relaxation time is sufficiently small, the numerical results obtained from Eq. (\ref{Sz_0}) deviate from the results obtained on the basis of the analytical formulas. These deviations are rather small for $\tau\gtrsim 10^{-11}$s, except the region near the zero chemical potential.  However,  the difference between the analytical and numerical results around $\mu =0$ is now much less pronounced than it was in the case of the $y$-component (compare Fig.2(e) and Fig.2(h). In turn, for $\tau\lesssim 10^{-11}$s the deviations become remarkable in the whole range of the chemical potentials shown in Fig.2(h).

All the components of the spin polarization ($S_x$, $S_{y}$ and $S_{z}$) vanish at $\mu = 0$ and are antisymmetric with respect to the sign reversal of the chemical potential. Numerical results presented above show that the spin polarization strongly depends on the Fermi level position.
In the close vicinity of the Dirac points, the $y$-component of the spin polarization has pronounced peaks (positive above and negative below $\mu =0$). The other two components behave more regularly in this region. All three components exhibit  some cusps (or dips) when $\mu$ is in the  vicinity of $\mu = \pm 2\lambda$, i.e.,  when the Fermi level approaches the top edge of the band $E_{1}$ or bottom edge of the band $E_{4}$. The spin polarization also remarkably depends on the Rashba parameter $\lambda$. This dependence reveals peculiarities of the corresponding electronic structure, and remarkably depends on the Rashba parameter.  In numerical calculations we assumed the Rashba spin-orbit coupling parameter $\lambda =2$ meV. Generally, this parameter depends on the substrate (or cover layer), and in real systems varies from a few to a few tens of meV, see eg. Refs [\onlinecite{Varykhalov,Marchenko12,Marchenko13,Balakrishnan2013,Balakrishnan2014,Avsar2014}].

\section{Vertex correction}

In the preceding section we have calculated spin polarization induced by electric field assuming effective relaxation time $\tau$ (or equivalently relaxation rate
$\Gamma$). Both, $\tau$ and chemical potential were treated there as independent parameters. When considering a specific relaxation mechanism, these parameters usually are not independent. Since the dominant scattering processes are  on impurities, we consider now this problem in more details.
Assume the scattering potential created by randomly distributed  weak short-range scatterers, which may be written  as $V(\mathbf{r}) s_{0} \sigma_{0} $ with Gaussian correlations $\langle V(\mathbf{r}) V(\mathbf{r}'))\rangle = n_{i} V^{2}\delta(\mathbf{r} - \mathbf{r}')$ (where $s_{0}$ and  $\sigma_{0}$ and  denote unit matrix in the pseudo-spin and spin subspace respectively).

Detailed calculation of the self energy due to scattering on the point-like impurities gives
$\Gamma_{1,4} = \frac{n_{i}V^{2}}{2v^{2}}(|\mu| - \lambda)$ and
$\Gamma_{2,3} = \frac{n_{i}V^{2}}{2v^{2}}(|\mu| + \lambda)$, where $n_i$ is the impurity concentration while $V$ is the impurity scattering potential. When $|\mu |\gg \lambda$, then indeed  $\Gamma_{1,4} \simeq \Gamma_{2,3}\equiv \Gamma$.
Otherwise, we take $\Gamma$ as the average of $\Gamma_{1,4}$ and $\Gamma_{2,3}$, i.e. $\Gamma = \frac{n_{i}V^{2}}{2v^{2}}|\mu|$.

When calculating the impurity averaged conductivity, it is well known that non-crossing diagrams give an important contribution and renormalize the results obtained in the {\it bare bubble} approximation. Such a vertex  renormalization is known to have a significant influence on the spin current induced {\it via} the spin Hall effect. In the case of two-dimensional electron gas with Rashba spin-orbit interaction it totally cancels the spin Hall conductivity obtained in the {\it bare bubble} approximation.\cite{Inoue04,Raimondi05,Dimitrova05,Chalaev05} However, this is not a general property and in other systems the vertex corrections can only reduce
partly the spin Hall effect.\cite{Murakami04,Malshukov05,Korotkov06}

The problem of disorder in graphene was discussed in many papers.~\cite{McCann2006,Ostrovsky2006,Pachoud2014} However, there is still a lack of information on the influence of disorder and impurities on spin-orbit driven phenomena in graphene. This problem was raised by Sinitsyn \textit{et al}.~\cite{Sinitsyn} and Gusynin \textit{et al.}~\cite{Gusynin2014} in the context of spin Hall and spin Nernst effect in the presence of intrinsic spin-orbit interaction in graphene and in the case of spin-independent random potential. In this case problem  becomes simpler because one can reduce the model to $2\times2$ space. Such a simplification, however, is not possible in the presence of Rashba spin-orbit interaction.

In the weak scattering limit, the localization corrections are vanishingly small and therefore only noncrossing ladder diagrams are important. The summation over the ladder diagrams can be  represented by the vertex corrections to the current-induced spin polarization. The renormalized spin vertex function is then given by the following equation:\cite{mahan}
 \begin{equation}
\tilde{S}_{\alpha} = \hat{S}_{\alpha} + n_{i} V^{2} \int \frac{d^{2} \mathbf{k}}{(2\pi)^{2}} G_{\mathbf{k}}^{A} \tilde{S}_{\alpha} G_{\mathbf{k}}^{R},
\end{equation}
where $\hat{S}_{\alpha}$ is defined by Eq.~(10).
For the point-like scattering potential  one can postulate the vertex function $\tilde{S}_{\alpha}$ in the form
\begin{eqnarray}
\tilde{S}_{\alpha} = a_{\alpha}\, \frac{\hbar}{2} \left(
                                    \begin{array}{cc}
                                      \sigma_{x} & 0 \\
                                      0 & \sigma_{x} \\
                                    \end{array}
                                  \right) + b_{\alpha}\, \frac{\hbar}{2} \left(
                                    \begin{array}{cc}
                                      \sigma_{y} & 0 \\
                                      0 & \sigma_{y} \\
                                    \end{array}
                                  \right)\nonumber\\ + c_{\alpha}\, \frac{\hbar}{2} \left(
                                    \begin{array}{cc}
                                      \sigma_{z} & 0 \\
                                      0 & \sigma_{z} \\
                                    \end{array}
                                  \right) + d_{\alpha}\, \frac{\hbar}{2} \left(
                                    \begin{array}{cc}
                                      \sigma_{0} & 0 \\
                                      0 & \sigma_{0} \\
                                    \end{array}
                                  \right),
\end{eqnarray}
for $\alpha =x,y,z$, where $a_\alpha$, $b_\alpha$, $c_\alpha$, and $d_\alpha$ are certain parameters to be determined. To find these parameters we multiply Eq.~(26) by the matrix as specified below and take the trace,
\begin{eqnarray}
{\mathrm{Tr}}\left\{\left(
                                    \begin{array}{cc}
                                      \sigma_{i} & 0 \\
                                      0 & \sigma_{i} \\
                                    \end{array}
                                  \right) \tilde{S}_{\alpha}\right\} = {\mathrm{Tr}}\left\{\left(
                                    \begin{array}{cc}
                                      \sigma_{i} & 0 \\
                                      0 & \sigma_{i} \\
                                    \end{array}
                                  \right) \hat{S}_{\alpha}\right\}\nonumber\\ + n_{i} V^{2} \int \frac{d^{2} \mathbf{k}}{(2\pi)^{2}} {\mathrm{Tr}}\left\{ \left(
                                    \begin{array}{cc}
                                      \sigma_{i} & 0 \\
                                      0 & \sigma_{i} \\
                                    \end{array}
                                  \right)   G_{\mathbf{k}}^{A} \tilde{S}_{\alpha} G_{\mathbf{k}}^{R} \right\},
\end{eqnarray}
for $i = 0,x,y,z$. Taking into account Eq.~(27), one finds then  a set of equations for the coefficients $a_{\alpha}, b_{\alpha}, c_{\alpha}, d_{\alpha}$.

We recall that in this paper the exchange field due to proximity effect is assumed to be small, so the current-induced spin polarization is limited to the terms linear in the exchange field. Consequently, the vertex correction is also calculated in the lowest order appropriate to have spin polarization linear in M.

For  $\alpha = x$  we find that:
\begin{eqnarray}
b_{x} = c_{x} = d_{x} = 0, \\
 a_{x} = \frac{1}{1 - n_{i} V^{2} \mathcal{I}_{x}},
\end{eqnarray}
where
\begin{equation}
\mathcal{I}_{x} = \int \frac{dk k}{2\pi}\frac{\chi_{x}(\mu, \Gamma)}{\prod_{n = 1}^{4} (\mu - E_{n} + i \Gamma)(\mu - E_{n} - i \Gamma)}
\end{equation}
and
\begin{eqnarray}
\chi_{x}(\mu, \Gamma) = k^{6} v^{6} + k^{4} v^{4} (3 \Gamma^{2}-\mu^{2})+(\Gamma^{2}+\mu^{2})^3 \nonumber \\
+4 (\Gamma^{4}-\mu^{4}) \lambda^{2}+k^{2} v^{2} (\Gamma^{2}+\mu^{2}) (3 \Gamma^{2}-\mu^{2}+4 \lambda^{2})\nonumber \\
\approx (k^{2} v^{2} -\mu^{2} )  \left(k^{4} v^{4}-\mu^{4}+4 \mu^{2} \lambda^{2}\right)\nonumber\\
 +\left(3 k^{4} v^{4}+3 \mu^{4} +2 k^{2} v^{2} \left(\mu^{2}+2 \lambda^{2}\right)\right)\Gamma^{2}.
\end{eqnarray}
In the above equation, only terms up to the second order in $\Gamma$ have been retained, while terms of higher order have been omitted.

For $\alpha = y$ we find the following coefficients:
\begin{eqnarray}
a_{y} = c_{y} = d_{y} = 0, \\
b_{y} = a_{x} = \eta.
\end{eqnarray}

In turn, for $\alpha = z$ we find
\begin{eqnarray}
a_{z} = b_{z} = d_{z} = 0, \\
c_{z} = \frac{1}{1 - n_{i} V^{2} \mathcal{I}_{z}} = \zeta ,
\end{eqnarray}
where
\begin{equation}
\mathcal{I}_{z} = \int \frac{dk k}{2\pi} \frac{\chi_{z}(\mu, \Gamma)}{\prod_{n = 1}^{4} (\mu - E_{n} + i \Gamma)(\mu - E_{n} - i \Gamma)}
\end{equation}
and
\begin{eqnarray}
\chi_{z}(\mu, \Gamma) = k^{6} v^{6} +k^{2} v^{2} \left(3 \Gamma^{2}-\mu^{2}\right) \left(\Gamma^{2} +
\mu^{2} \right)\nonumber\\
+ \left(\Gamma^{2}+ \mu^{2}\right)^{3} +k^{4} v^{4} \left(3 \Gamma^{2}-\mu^{2} - 2 \lambda^{2}\right)\nonumber\\+\left(\Gamma^{2}+ \mu^{2}\right) \left(2 \left(\Gamma^{2} -3 \mu^{2}\right) \lambda^{2} + 8 \lambda^{4} \right)\nonumber \\
\approx \left((k^{2} v^{2}-\mu^{2})^{2} - 4 \mu^{2} \lambda^{2} )\right) \left(k^{2} v^{2} + \mu^{2} - 2 \lambda^{2} \right)  \nonumber\\
+ \left(3 k^{4} v^{4} + 2 k^{2} v^{2} \mu^{2} + 3 \mu^{4} - 4 \mu^{2} \lambda^{2} + 8 \lambda^{4} \right) \Gamma^{2}.
\end{eqnarray}

Finally the renormalized spin-vertex functions are:
\begin{equation}
\tilde{S}_{x} = \frac{\hbar}{2} \eta \left(
                                    \begin{array}{cc}
                                      \sigma_{x} & 0 \\
                                      0 & \sigma_{x} \\
                                    \end{array}
                                  \right)
\end{equation}
\begin{equation}
\tilde{S}_{y} = \frac{\hbar}{2} \eta \left(
                                    \begin{array}{cc}
                                      \sigma_{y} & 0 \\
                                      0 & \sigma_{y} \\
                                    \end{array}
                                  \right)
\end{equation}
\begin{equation}
\tilde{S}_{z} = \frac{\hbar}{2} \zeta \left(
                                    \begin{array}{cc}
                                      \sigma_{z} & 0 \\
                                      0 & \sigma_{z} \\
                                    \end{array}
                                  \right)
\end{equation}

This means that the results obtained in the {\it bare bubble} approximation should be multiplied only by a numerical factor to take into account the vertex corrections due to disorder. More specifically, the results for $S_{x}$ and $S_{y}$ should be multiplied by the factor $\eta$ while those for $S_{z}$ should be multiplied by $\zeta$. The situation is significantly different from that found in the case of spin Hall effect. This is because  transport phenomena and spin polarization are affected by scattering on impurities in remarkably different ways.

\begin{figure*}[t]
 \centering \includegraphics[width=0.9\textwidth]{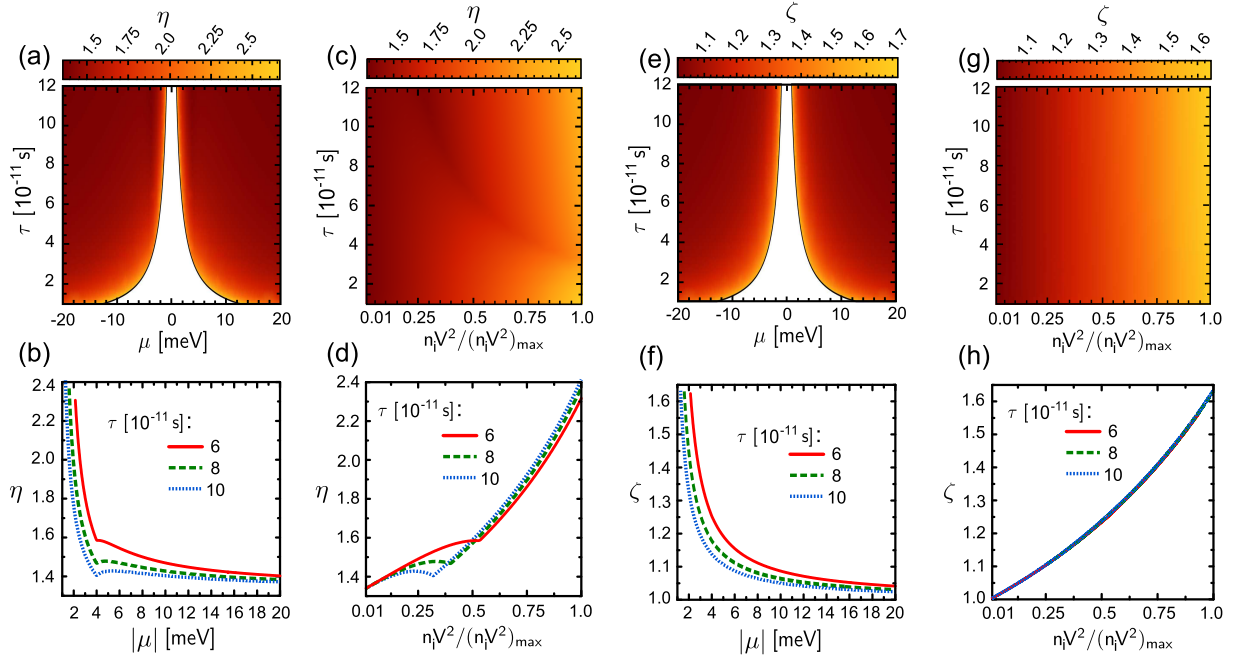}
\caption{(Color online) The parameter $\eta$ as a function of the chemical potential and relaxation time (a) and as a function of relaxation time and ${n_{i}V^{2}}/({n_{i}V^{2}})_{\rm max}$ (c).  Figures (b) and (d) show  $\eta$  as a function of chemical potential $\mu$ (b)  and ${n_{i}V^{2}}/({n_{i}V^{2}})_{\rm max}$ (d) for indicated values the relaxation time.
Figures (e)-(h) show the same variations as figures (a)-(d), but  for the parameter $\zeta$.
The white regions in (a) and (e) are excluded for the assumed value of
$({n_{i}V^{2}})_{\rm max}=0.4\times 10^{-2} ({\rm eV}\cdot {\rm nm})^2$.  The other parameters are as in Fig.2.}
\end{figure*}

In Fig.3(a) we show the renormalization parameter $\eta$ as a function of chemical potential and relaxation time. Now the relaxation time is connected with the chemical potential through the relation $\frac{\hbar}{\tau} = \frac{n_{i}V^{2}}{v^{2}}|\mu|$. A single point in the $\tau ,\mu$ space corresponds to a well defined value  of $n_{i}V^{2}$. However, possible values of $n_{i}V^{2}$ have been limited in Fig.3 to $n_{i}V^{2}<({n_{i}V^{2}})_{\rm max}$, where
$({n_{i}V^{2}})_{\rm max}$ is a certain maximum value which is physically reasonable.
The central white region is bounded by the condition $\hbar /\tau = (n_{i}V^{2})_{\rm max}|\mu|/v^{2}$ and is excluded for the considered parameters. In Fig.3c, in turn, we show the parameter $\eta$ as a function of  the relaxation time and the ratio ${n_{i}V^{2}}/({n_{i}V^{2}})_{\rm max}$.
As one might expect, this figure shows that the normalization parameter $\eta$ becomes reduced with decreasing  ${n_{i}V^{2}}$.
Figures 3(b) and 3(d) present cross-sections of Fig.3(a) and 3(c), respectively. The above described results for $\eta$  show that the $S_x$ and $S_y$ components  are remarkably renormalized by the vertex correction and are enhanced by a factor of the order of 2 (between  1 and 3).  This enhancement of the spin polarization is comparable to that found in the case of two-dimensional electron gas with Rashbe interaction.\cite{edelstein90}
The parameter $\zeta$, in turn, is shown in Fig.3e-f. It is of the same order of magnitude as the parameter $\eta$ and depends on the chemical potential and relaxation time in a similar way, so we will not discuss it in more detail.

\section{Spin-orbit torque}

The current-induced spin polarization is exchange-coupled to the local magnetization $\bf M$ and thus exerts a torque on $\bf M$.
According to Eq.(\ref{hkM}), energy of this interaction per unit area  can be written as $E_{\rm ex} = -(2/\hbar )\tilde{\bf M}\cdot \bf{S}$, where $\bf{S}$ is the induced spin polarization. Taking into account the relation between $\tilde{\bf M}$ and $\bf M$, one finds the spin-orbit torque per unit area, $\btau$, exerted on the magnetization (more precisely on the corresponding equilibrium spin polarization of the system) in the form
\begin{equation}
\label{sot}
\btau = \frac{2}{\hbar} \tilde{\mathbf{M}} \times \mathbf{S}= \frac{J_{\rm ex}}{g\mu_B\hbar} \mathbf{M} \times \mathbf{S}.
\end{equation}

Let us consider in more detail some specific situations as concerns relative orientation of the magnetization and electric field (current). Let us start with the situation when the magnetization $\mathbf{M}$ is in the plane of the system and perpendicular to the current. This corresponds to $\theta = 0$ and $\xi = 0$ ($M_{x} = M \neq 0$ and $M_{y} = M_{z} = 0$). From the above general equation follows that the spin-orbit torque can be then written in a general form as
\begin{equation}
\label{sot1}
\btau =A( - \hat{j} M_{x} S_{z} + \hat{k} M_{x} S_{y}),
\end{equation}
where $\hat{i}$, $\hat{j}$, and $\hat{k}$ are unit vectors along the axes $x$, $y$ and $z$, respectively, and we introduced the following abbreviation: $A= J_{\rm ex}/g\mu_B\hbar$.
Taking into account  Eqs (\ref{sy1}), (\ref{sy2}), (\ref{sz1}) and (\ref{sz2}), one finds immediately that
the spin-orbit torque in this geometry disappears because  both $S_{y}$ and $S_{z}$ component of the spin polarization vanish.

Consider now the situation corresponding to $\theta = 0$ and $\xi = \pi/2$ ($M_{y} = M \neq 0$ and $M_{x} = M_{z} = 0$), i.e. the case when the magnetization is parallel to the electric current. From Eq.(\ref{sot}) follows that the spin-orbit torque has the following general form:
\begin{equation}
\label{sot2}
\btau =A( \hat{i} M_{y} S_{z} - \hat{k} M_{y} S_{x}).
\end{equation}
The $S_{z}$ component is now nonzero, and thus both, $S_z$ and $S_x$ contribute to the torque in this geometry.

When $\theta = \pi/2$, namely $M_{z}=M \neq 0$ and $M_{x} = M_{y} = 0$, the magnetization is perpendicular to the graphene plane. The spin-orbit torque takes then the  general form,
\begin{equation}
\label{sot3}
\btau =A( - \hat{i} M_{z} S_{y} + \hat{j} M_{z} S_{x}).
\end{equation}
Similarly as in the preceding situation, both $S_y$ and $S_x$ are nonzero and determine the torque.

In the last two cases the spin-orbit torque contains two components: linear term with respect to $J_{\rm{ex}}$ (proportional to $S_{x}$) and quadratic term in $J_{\rm{ex}}$  (proportional to $S_{z}$ and $S_{y}$).
The spin orbit torque contains one component proportional to the relaxation time and another component whose the dominant part is independent on the relaxation time.

In a general case of arbitrary orientation of the magnetic moment, magnitude and character of the spin-orbit torque varies with the orientation of the magnetic moment. This is because two components of the current-induced spin polarization depend on the magnetization, while the third one is independent of $\bf M$.  As a result the  spin torque may have field-like and (anti)damping terms.

\section{Summary}
We have calculated current-induced spin polarization in graphene deposited on a ferromagnetic substrate, that ensures not only Rashba spin-orbit interaction but also a ferromagnetic moment in the graphene layer. To describe electronic spectrum of graphene we have used Kane Hamiltonian that describes low-energy states around the Dirac points. Using the zero-temperature Green functions formalism and linear response theory, we have derived analytical formulas for the spin polarization, up to the terms linear in $M$. Numerical results based on the analytical formulas have been compared with those obtained by numerical integration procedure. From this comparison we have formulated applicability conditions of the analytical results. Significant deviations of the analytical results from those based on numerical integration have been found for relaxation times smaller than $10^{-10}$s.

The nonequilibrium (current-induced) spin polarization exerts a torque on the magnetization {\it via} the exchange interaction. This torque contains a term which is proportional to the $x$-component of the induced spin polarization and therefore is proportional to the momentum relaxation time.  The torque  also includes a component whose main part is independent of the relaxation time.

The spin-orbit torque due to the interplay of external electric field and Rashba coupling at the interface between graphene and a magnetic layer can be used  for instance to trigger magnetic dynamics and/or magnetic switching. Indeed, such a switching  was observed experimentally in a recent paper by Wang {\it et al}.\cite{wangpreprint}  However, instead of graphene they used MoS$_2$ --  another two-dimensional honeycomb crystal.

\begin{acknowledgments}
This work has been partially supported by the National Science Center in Poland as research
project No. DEC-2013/10/M/ST3/00488 and by the Polish Ministry of Science and Higher Education
through a research project 'Iuventus Plus' in years 2015-2017 (project No. 0083/IP3/2015/73).
\end{acknowledgments}

%%%%%%%%%%%%%%%%%%%%%%%%%%%%%%%%%%%%%%%%%%%%%%%%%%%%%%%%%%%%%%%%%%%%%%%%%%%%%%%%%%%%%%%%%%%%%%%%%%%%%%%%%

\end{document}